\documentclass[aps,twocolumn,showpacs]{revtex4-1}
\usepackage{graphicx}

\begin{document}
\title{Application of the Wang--Landau method to  faceting phase transition.}
\author{  C. Oleksy}
\affiliation{Institute of Theoretical Physics, University of
Wroc{\l}aw,
 Plac Maksa Borna 9, 50-204 Wroc{\l}aw, Poland
}
\date{\today}
\begin{abstract}
A simple solid--on--solid model of adsorbate -- induced  faceting
is studied by using a modified Wang--Landau method. The phase
diagram for this system is constructed by computing the density of
states in a special two--dimensional energy space. A finite--size
scaling  analysis of transition temperature and specific heat
shows that faceting transition is the first order phase
transition. Logarithmic dependence of the mean--square width of
the surface on system size indicates that  surface is rough above
the transition temperature.

\end{abstract}

\pacs%
{%
 68.35.Rh,
 68.43.De,
 64.60.Cn
}

\maketitle

\section{Introduction}

It has been recently demonstrated
\cite{song95,mad96,mad99a,mad99b,song07} that
surfaces such as  W(111) and Mo(111) covered with a single
physical monolayer of certain metals (Pd, Rh, Ir, Pt, Au) undergo
massive reconstruction from a planar morphology to a  faceted
surface upon annealing at $T>700$K. The faceted surface is covered
by three--sided pyramids with mainly \{211\} facets. A major
mechanism for facets formation is minimization  of total surface
free energy\cite{leung97}.  Faceting of bcc(111) and
fcc(210) surfaces  can also be induced by oxygen or  other
nonmetallic impurities (see  Ref.\cite{mad08} and references therein).
Adsorbate -- induced faceting is also observed  on curved surfaces
\cite{tsong01,szczep05,szczep05b} and this
phenomenon has been recently used in fabrication of electron and
ion point sources \cite{tsong08,tsong09}.

Investigations of thermal stability of faceted surfaces have
revealed that reversible phase  transition occurs in  systems
Pd/Mo(111)\cite{song95},  O/Mo(111)\cite{song95}, and
Pd/W(111)\cite{song07}. As a result of faceting transition, the
faceted surface changes into planar surface at the transition
temperature.

In theoretical studies of complicated surface problems (e.g.
roughening transition, surface reconstruction, surface growth,
surface phase transitions)   simple models  like lattice gas
models or  solid--on--solid (SOS) models are applied
\cite{beijeren95,tosat96,jaszczak97,jaszczak97ss,kaseno97}.
In our earlier paper \cite{oleksy} we have proposed a  SOS model
to study the adsorbate--induced faceting of the bcc(111) crystal
surface at constant coverage. Monte Carlo simulation results show
formation of pyramidal facets in accordance with experimental
observations. Moreover, the model describes a reversible phase
transition from a faceted surface to a disordered (111) surface.
This model reproduces also  formation of \{211\} step--like facets
on curved surfaces \cite{szczep05b,Niewiecz06}.

Simulation results \cite{oleksy} -- based on the Metropolis
algorithm, indicate that the faceting  is the first--order
transition. However, identification of the nature of this
transition in finite--sized system  is difficult and it can be
solved by use of finite--size scaling\cite{kosterlitz91}. Another
way to identify the first--order transition is to study the
distribution of energy which has a double--peak structure in the
vicinity of transition temperature. This can be  easily
accomplished by use of the  Wang -- Landau (WL)
method\cite{wl_0,wl_1}. The WL method is based on accurate
calculation of density of states and therefore allows for
calculating thermodynamical functions. The  method is especially
useful in investigation of phase transitions \cite{wl_0,wl_1,wl_3,stampfl}
to determine the order of transition, transition temperature and behavior of
thermodynamical quantities.

In this paper we use WL method to study faceting transition in
overlayer--induced faceting on bcc(111) surface. A short
presentation of SOS model of adsorbate -- induced faceting is
given in Sec.~\ref{s2}. Application of  WL method and its
modification proposed by  Belardinelli and Pereyra\cite{pereyra_1}
to calculation density of states is presented in  Sec.~\ref{WL}.
Sec.~\ref{ScPD} contains a method of construction of a phase
diagram for a finite system with competitive interactions. In this
method, energy of the system  is decomposed  into a few parts and
density of states is calculated  in multi--dimensional energy
space. Due to this method, the phase diagram is constructed for
the SOS model of finite sizes. In order to determine the order of
phase transition, the finite -- size scaling of the transition
temperature and specific heat is presented in Sec.~\ref{ScFS}. The
nature of high--temperature phase is investigated in
Sec.~\ref{ScHTP}. Contrary to experimental results, the finite
size analysis of  the mean--square width of the surface indicates
that above the transition temperature a surface is rough.

\section{The SOS model}\label{s2}

\begin{figure}
\centering
\includegraphics[width=8cm]{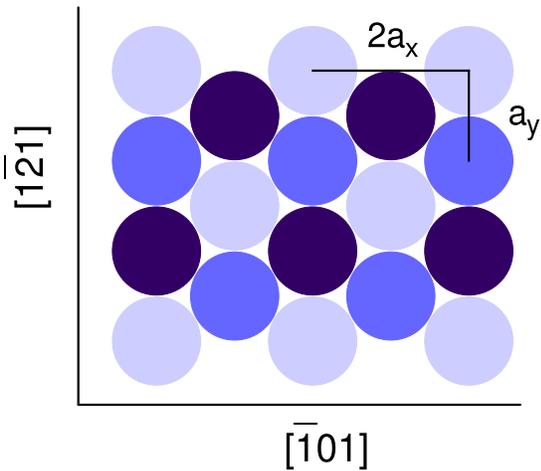}
\caption{ Schematic top view of the bcc(111) surface. Atoms from
three successive geometrical layers represent positions of columns
in the SOS model. The Z axis is normal to the (111) plane and
$a_x=a\sqrt{2}/2$, $a_y=a\sqrt{6}/3$.} \label{bcc111}
\end{figure}

To study  an adsorbate--induced faceting on bcc(111) surface a
solid--on--solid model has been proposed\cite{oleksy}. The model
consist of columns placed on the triangular lattice obtained by
projection of the bcc crystal lattice on the (111) plane (see
Fig.~\ref{bcc111}). A column height  $h_i$ at site $i=(i_x,i_y)$
in the {\em l}th sublattice, $l=0, 1, 2$   takes discrete values
of the form $h_i=3 n_i +l$ where  $n_i$ is the number of atoms in
this column.
The model is designed for constant coverage of 1 physical
monolayer -- critical coverage for adsorbate--induced faceting
\cite{mad99b}. This means that each column has exactly  one
adsorbate atom placed at the highest position. There is also a
restriction imposed on column heights: the nearest neighbor column
heights can differ only by $\pm 1, \pm 2$.

The  surface formation energy  can be expressed as interaction
energy  of columns
\begin{equation}\label{hsos}
\begin{array}{lll}
{\cal H} &=&
  \frac{1}{2}\sum\limits_i
    \left\{
      \sum\limits_{j_1}
         {\left[
           J_1 \omega_{i,j_1 }(1)
             + K_1 \omega_{i,j_1 }(2)
             \right]
         }  \right.\\[3ex]
   & &
     + \sum\limits_{j_2}
           {\left[
              2 J_2 \omega_{i,j_2 }(0)
               + \left( 2 J_2  + K_2\right)
               \omega_{i,j_2 }(3)

           \right] }  \\
   & &
     +
     \left.
       J_2 \sum\limits_{j_3}
         {\left[
                \omega_{i,j_3 }(2)

               + \omega_{i,j_3 }(4)

         \right]}
     \right\}+ N J_0,
\end{array}
\end{equation}
where the sums over $j_1$, $j_2$, and $j_3$ represent the sums
over first, second, and third neighbours of the column at site
$i$, respectively. The function  $\omega_{i,j}(k)= \delta\left(
\left| h_i - h_j \right| - k \right)$, for $k=0,1,2,\ldots$, is
expressed by the Kronecker delta: $\delta(x)=1$ for $x=0$ and
$\delta(x)=0$ for $x\neq0$. Model parameters $J_0$, $J_1$,  $J_2$,
$K_1$, and $K_2$ depend on interaction energies between substrate
and adsorbate atoms (for details see \cite{oleksy}).

It turns out that  the energy of the nearest neighbours interaction
in the Hamiltonian is  conserved. This property
follows from symmetry of the model and the restriction imposed on
the on column heights. Thus the energy of the nearest neighbours
interactions  can be treated as the reference energy. Choosing
coupling constant $J_2$ as the unit of energy, we will work with
one  model parameter $K = K_2/J_2$. The dimensionless Hamiltonian
$\tilde{\cal H} = {\cal H}/J_2 $, takes the following form
\begin{equation}\label{hsoseff}
\begin{array}{lll}
\tilde{\cal H} &=&
  \frac{1}{2}\sum\limits_i
\left\{
      \sum\limits_{j_2}
           {\left[
              2 \omega_{i,j_2 }(0) + \left( 2+ K\right) \omega_{i,j_2 }(3)

           \right] }  \right.\\[3ex]
   & &
     +
     \left.
        \sum\limits_{j_3}
         {\left[
               \omega_{i,j_3 }(2)+ \omega_{i,j_3 }(4)
         \right]}
     \right\}+ \tilde {\cal H}_{\mathrm{ref}},
\end{array}
\end{equation}
 In what follows we will use dimensionless energy omitting  the tilde and
$\tilde{\cal H}_{\mathrm{ref}}$. It has been shown \cite{oleksy}
that the energy of the (211) face is minimal when $-2<K<0$,
whereas the (111) surface is stable for $K>0$. For $K<-2$ the
(011) face  has minimal energy.

\section{Density of states}\label{WL}

In this paper we calculate density of state for the SOS model
using a modification of the Wang--Landau  method \cite{wl_1}
proposed by Belardinelli and Pereyra\cite{pereyra_1,pereyra_2}.

\subsection{The Wang -- Landau method}
The WL method is based on a random walk which produces a flat
histogram in the energy space. A trial configuration of energy
$E^{\prime}$ is accepted with probability
\begin{equation}\label{trans_probab}
W(E\rightarrow E^{\prime})=min[1,g(E)/g(E^{\prime})],
\end{equation}
 where E
is energy of the current configuration and $g(E)$ means the
density of states. However, the $g(E)$ is not known and
calculation of the density of states is the main goal of WL
method. To achieve this goal, the $g(E)$ is changed after each
step of the random walk $g(E)\rightarrow fg(E)$ by  a modification
factor $f>1$. Moreover, it is assumed that  initially $g(E)=1$ for
all energies and $f=e$.

Very recently, Belardinelli and Pereyra have demonstrated
\cite{pereyra_1} saturation of errors in the WL method, or
nonconvergence  of calculated density of states to the exact
value. Moreover, they have shown that if the refinement parameter
$\ln f(t)$ depends on time as  $\propto t^{-1}$ for large time,
the calculated density of states approaches asymptotically to the
exact values as   $\propto t^{-1/2}$. This fact is used in
Belardinelli -- Pereyra (BP) modification of the WL method.

To present the BP method  let us introduce some quantities. It is
assumed that the random walk is performed in the energy range
$E_{min}\leq E \leq E_{max}$ with $\Omega$ different energy
levels.  A Monte Carlo  time  $t=j/\Omega$ is defined as the number
of trial configurations \emph{j}  used so far
with respect to the number of energy states $\Omega$. From
numerical reason it is convenient to use $S(E)={\ln g(E)}$ and
$F=\ln f$ instead of $ g(E)$ and $f$.
There are two stages in calculation of $ S(E)$ in the BP method.
In the first stage, the refinement parameter $F$ is changed
similarly as in the original WL method, i.e., $F \rightarrow F/2$
when all energy states are visited in the random walk with given
F. Please notice that the criterion for flatness of the histogram
is not used here. The second stage begins at critical time $t_c$
defined as a moment when the new value of $F$ becomes smaller than
$1/t$. From this time, the refinement parameter takes the
continuous form $ F(t)= t^{-1}$. The second stage lasts   until
the refinement parameter reaches a predefined value
$F_\mathrm{{min}}$ (typically $F_\mathrm{{min}} =10^{-8}$).
To control the convergence of S(E,t) during the  random walk it is
useful to measure  the following quantities: the histogram
$H(E,t)$ and its  averaged value  at time t, $\langle H(t)\rangle=
\frac{1}{\Omega}\sum_E{ H(E,t)}$, and the width of the histogram
${\Delta H(t)= H_\mathrm{{max}}(t)- H_\mathrm{{min}}(t)}$, and the
relative width $\delta H(t)=\frac{\Delta H(t)}{\langle
H(t)\rangle}$. According to result of Ref.\cite{pereyra_2} the
quantity $\delta H(t)$ has the same longtime behaviour as the
error of $S(E,t)$
\begin{equation}
 \delta H(t)\propto  t^{-\frac{1}{2}}
\end{equation}
As the exact value of $S(E)$ is not known, the time dependence of
$\delta H(t)$  can be used to evaluate   convergence and the
accuracy of BP algorithm.

\subsection{Calculation of S(E) for the SOS model}

 We consider the SOS model on the  rectangular lattice with $N_x$
and $N_y$ columns along x and y axis, respectively, and with
periodic boundary conditions. A relation    $N_x =\frac{7}{6}N_y$
is assumed to assure approximate equality of linear lattice sizes
along the x and y axis.  The linear system size is defined as
$L=\sqrt{N_xN_y}$. The quantity $S(E)=\ln g(E)$ is calculated by
performing the the random walk in the energy space with the
transition probability given by Eq.~(\ref{trans_probab}). At each
step a trial configuration is generated by choosing two lattice
sites \emph{i} and \emph{j} and changing heights of columns at
these sites: $(h_i, h_j)\rightarrow (h_i-3, h_j+3)$. Due to
constrains imposed on columns height in the SOS model, the change
$(h_i, h_j)\rightarrow (h_i-3, h_j+3)$ is allowed only when $h_i$
is a local maximum and $h_j$ is a local minimum. Local maximum
(minimum) at site $k$ denotes that column $h_k$ is higher (lower)
than its 6 nearest neighbor columns, respectively. In order to
speed up calculation we use lists of local maxima and minima and a
trial configuration is generated by random choice of a maximum and
a minimum.

During the preliminary application of the BP method to the SOS
model we encountered a problem of very large $t_c$ even for the
small linear sizes of the system. Contrary to simulation of Ising
model,  the number of steps  needed to visit each energy level at
least once becomes very large even for the first value of the
refinement parameter $F_{0}=1$.  To overcome this problem, we modify
the first stage of the BP method by introducing a separation $s$ between
successive increments of S(E) and H(E) in the random walk,
similarly as in Ref.\cite{bhatt}.

The errors $\delta S(E)$ of $S(E)$ are estimated from sample of
$m$ independent $S_1(E), \ldots S_m(E)$ results of simulations and
then  an average  error $\delta S= \frac{\sum_E{ \delta
S(E)}}{\Omega -1}$ is calculated. Having calculated $S(E)$ one can
easily investigate temperature dependence of  various quantities
discussed in the paper:

the energy distribution
\begin{equation}\label{E_probab}
P(E,T)=\frac{ \exp(S(E) -E/T)}{\sum\limits_{E} \exp(S(E)-E/T) },
\end{equation}

moments of energy
\begin{equation}
\langle E^n \rangle=\frac{\sum\limits_{E}E^n\exp(S(E)-E/T)}
{\sum\limits_{E}\exp(S(E)-E/T) },
\end{equation}

the specific heat per site
\begin{equation}\label{spec_heat}
C=\frac{\langle E^2\rangle -\langle E\rangle^2}{L^2T^2},
\end{equation}

the Binder's fourth cumulant

\begin{equation}\label{Binder_cum}
V_4=1-\frac{\langle E^4\rangle }{3\langle E^2\rangle^2}.
\end{equation}

\section{Phase Diagram}\label{ScPD}

WL and BP methods can be easily applied to construct a phase
diagram for a finite system with competitive interactions by
performing the random walk in multi--dimensional energy space. In
this paper we construct the phase diagram in ($T$,$K$) plane by
use of a modified BP method and performing the random walk in
two--dimensional energy space ($E_J, E_K$). To do this the energy
of the system, $E=\tilde{\cal H} - \tilde {\cal H}_{\mathrm{ref}}$
from Eq.~(\ref{hsoseff}), is decomposed into two parts  $E=E_J +K
E_K$, where $K E_K$ represents  interaction energy with the
coupling constant $K$ and $E_J$ stands for remaining contribution
to $E$.

Having calculated density of states $g(E_J, E_K)$ or $S(E_J,
E_K)=\ln g(E_J, E_K)$ one can easily obtain the mean energy
$\langle E\rangle$, the specific heat $C$, and other quantities
for any value of the coupling constant $K$. For example, the mean
energy can be calculated as

\begin{equation}
\langle E(K)\rangle=\frac{ \sum\limits_{E_J}\sum\limits_{E_K}(E_J
+K E_K ) e^{S(E_J, E_K)-\frac{E_J +K E_K}{T}}
}{\sum\limits_{E_J}\sum\limits_{E_K}e^{S(E_J, E_K)-\frac{E_J +K
E_K}{T}} }
\end{equation}
This approach is limited to rather small systems because the
number of states  $\Omega(E_J, E_K)$  is much more greater than
$\Omega(E)$ in one--dimensional energy space. For example,
$\Omega(E_J, E_K)$  reaches 82073 and 464162 for $L=26$ and $39$,
respectively whereas $\Omega(E)$ amounts to 304 and 718 for
$K=-1$. Therefore, we limit study of the phase diagram to
two system sizes,  $L=26$ and $39$.

In simulations of $S(E_J, E_K)$ we used separation $s=4L$ and the
refinement parameter $F$ was limited by $F_\mathrm{{min}}
=10^{-7}$. For such parameters and system size $L=39$ the
computation takes about 15 days on a 2.6 GHz Opteron processor.
\begin{figure}
    \includegraphics[width=8cm]{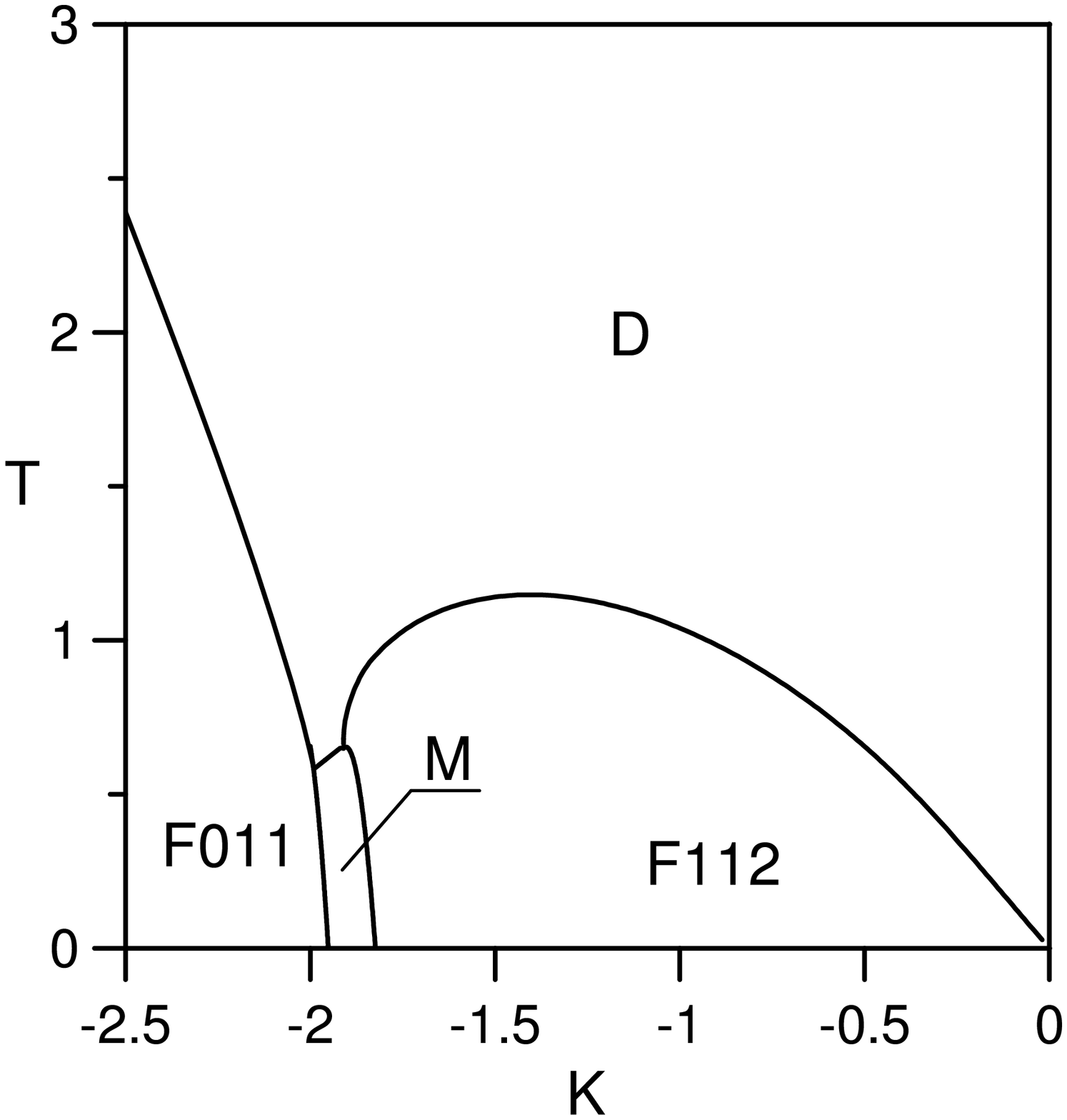}
    \caption{
 Phase diagram for system size $L=39$. Labels D, F112, F011, and M
 denotes disordered, faceted \{112\}, faceted \{011\}, and mixed phases, respectively.
    }
    \label{Diagram}
\end{figure}
On the other hand, the average error of $S(E_J, E_K)$ estimated on
results of four independent runs, was rather small: $3*10^{-3}$
and $2*10^{-3}$ for $L=26$ and $L=39$, respectively. Hence, this
validates the use of $S(E_J, E_K)$ to calculate thermodynamical
quantities.

To construct the phase  diagram in the ($T,K$) plane for a finite
system of a linear size $L$ we treat temperature $T_c(L,K)$ at
which the specific heat has a maximum, as temperature  of a phase
transition. A simple optimization method -- golden section search
\cite{numrec} was applied  to locate the maximum of the specific
heat as a function of temperature for a given $K$. We also
examined temperature dependence of the Binder's fourth cumulant
Eq.~(\ref{Binder_cum}) because $V_4$ has a minimum at a
first--order phase transition.

\begin{figure}
    \includegraphics[width=8cm]{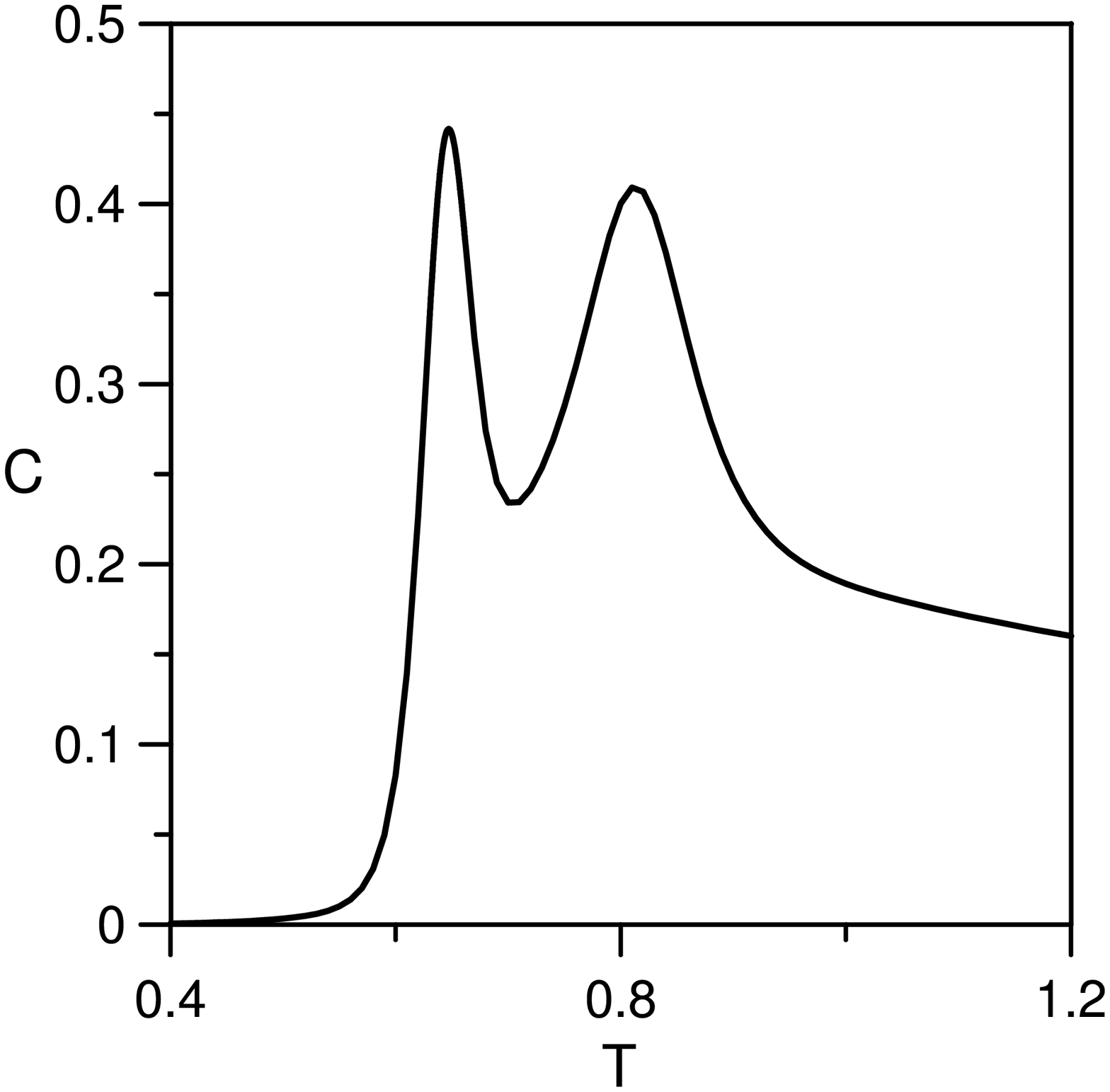}
    \caption{
 Specific heat with two peaks for $K=-1.90$.
    }
    \label{2peak}
\end{figure}

\begin{figure}
    \includegraphics[width=8cm]{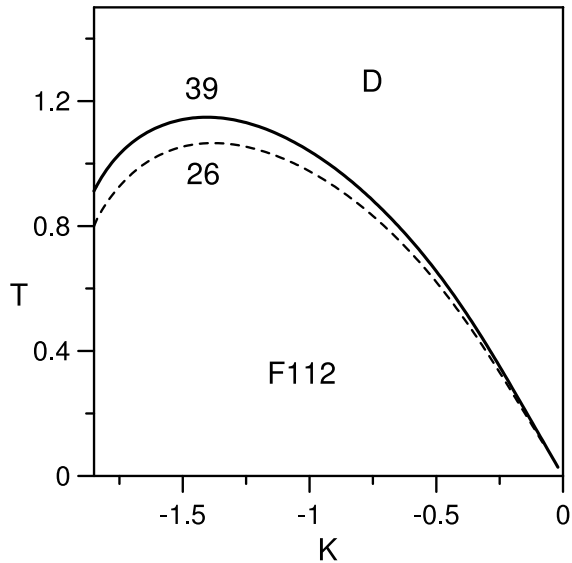}
    \caption{
 Finite--size effect. The transition line between the faceted  F112  and  disordered D phases.
 Continuous (dashed) line represent system size $L$ 39 (26), respectively.
    }
    \label{Diag_fse}
\end{figure}

\begin{figure}
    \includegraphics[width=8cm]{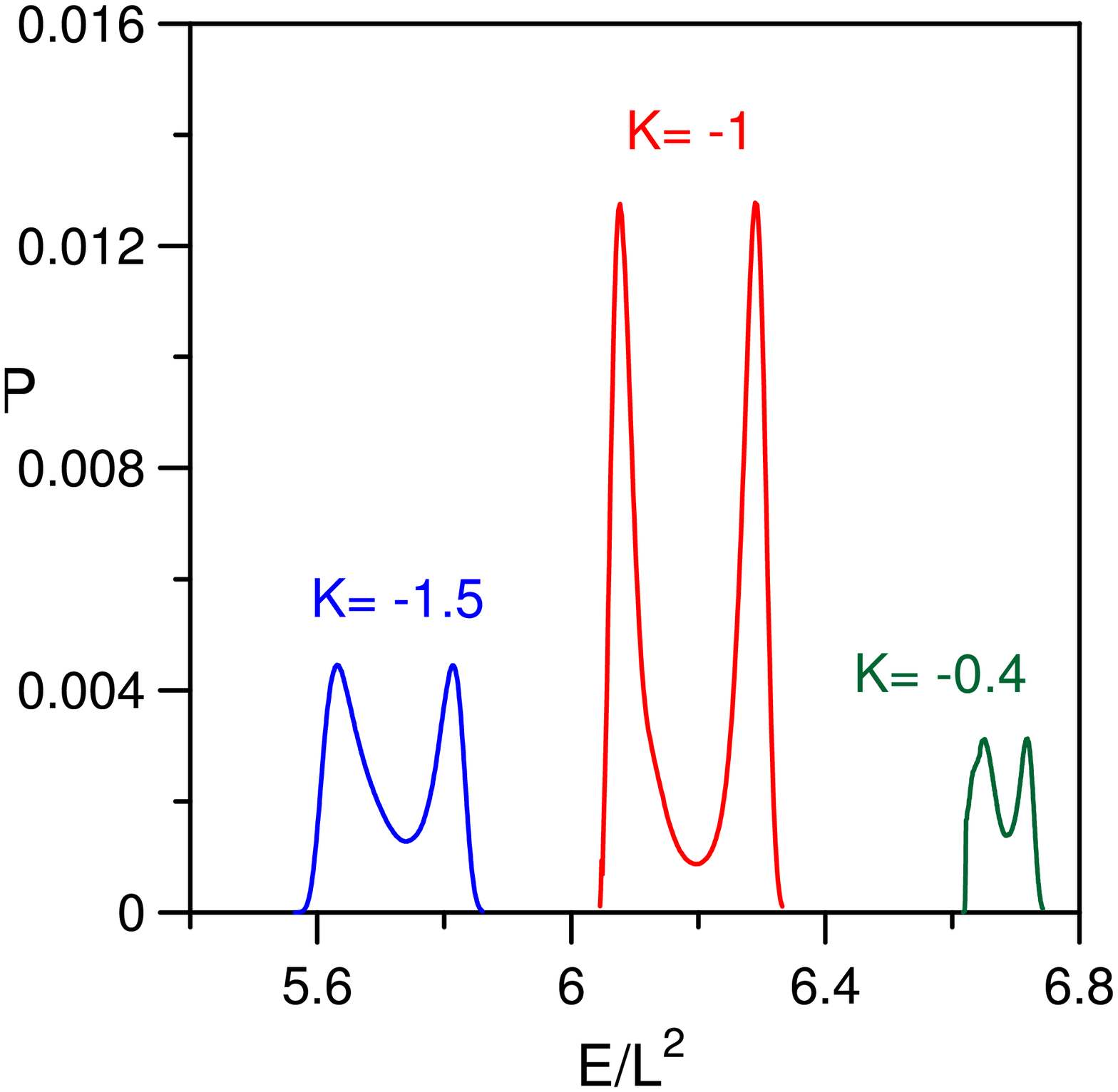}
    \caption{
Energy distribution at transition temperature for three values of
$K$ and  system size $L=39$.
    }
    \label{pe_diag}
\end{figure}

The phase diagram (see Fig.~\ref{Diagram}) constructed for the
lattice size $L=39$ comprises three phases: faceted {112}, faceted
{011}, and disordered. There is also a region between two faceted
phases where the mixture of these phases appears. The line
separated a faceted phase and the mixed one is determined from
location of the additional maximum in the specific heat (see for
example Fig.~\ref{2peak}). Thus, in this case the first peak
corresponds to transition from faceted $112$ phase to the mix one,
whereas the second peak is generated by transition from faceted
$112$ phase to the disordered phase.
The phase diagram for a smaller linear size $L=26$ has the same
qualitative form as the diagram in Fig.~\ref{Diagram}  but the
phase transition lines are slightly shifted  due to the
finite--size effects. The largest differences are observed for
transition temperature from the $F112$ phase to the disordered one
(see Fig.~\ref{Diag_fse}). In what follows  we limit  our
consideration to this transition because it is observed
experimentally \cite{song95,song07}. Big differences between
transition temperatures for these small systems require the
finite--size analysis for larger linear system sizes. This can be
performed by calculation of density of states in one--dimensional
energy space. Inspection of energy distribution indicates that
transition between the $F112$ and the disordered phase is of first
order phase transition because  the energy distribution at
transition temperature has the double peak form (see
Fig.~\ref{pe_diag}). The first peak represents the $F112$ phase
and the second peak represents the disordered phase.  The nature
of this phase transition is also confirmed by  temperature
dependence of the Binder's fourth cumulant $V_4$, which has
minimum at temperature close to $T_c(L)$. Thus, it is enough to
perform a detailed analysis for one  value of the coupling
constant $K$, and we choose the case $K=-1$ to minimize the number
of energy states or computing time.

\section{Finite--size scaling}\label{ScFS}

The $S(E)$ is calculated for  the coupling constant $K=-1$  in  a single  energy
interval $(E_{\mathrm{min}}, E_{\mathrm{max}})$ in order to avoid
boundaries errors caused by partition of energy space into several
pieces. It is especially important in a case of first order phase
transition, where at transition temperature $T_c$, the energy
distribution Eq.~( \ref{E_probab}) has two peaks of equal heights
at $E=E_{-}$ and $E=E_{+}$.  In our case the interval $(E_{-},
E_{+})$ makes up about $1/3$ of the whole energy interval.  The
energy of planar face (111) is chosen as $E_{\mathrm{max}}$ and
this state has known degeneracy $g(E_{\mathrm{max}})=1$. On the
other hand, it is difficult to reach states close to the minimal
energy due to edge energies, hence we choose $E_{\mathrm{min}}$ as
small as possible to assure convergence of $S(E)$.

To study the size dependence of some physical quantities the
following numbers $L$  where used in calculations 26, 39, 52, 65,
78, 91, and 104. For each value of $L$ the $S(E)$ was calculated
until refinement parameter $F$ reached the minimal value
$F_\mathrm{{min}} =10^{-8}$. Values of separation $s$ used in the
first stage of BP method were comparable with the number of energy
states ($s\approx 0.7\Omega$). To estimate errors of $S(E)$ each
calculation were repeated 5 times and average density was
obtained. The relative averaged error of $S(E)$ was smaller than
$3\times 10^{-4}$ for $L <104$ and $1.4\times 10^{-3}$ for $L
=104$. The increase of the error of $S(E)$  for the largest size
studied here was caused by large fluctuation of the histogram in
the low energy range. Therefore, we did not study the systems with
sizes $L>104$.

\begin{figure}
    \includegraphics[width=8cm]{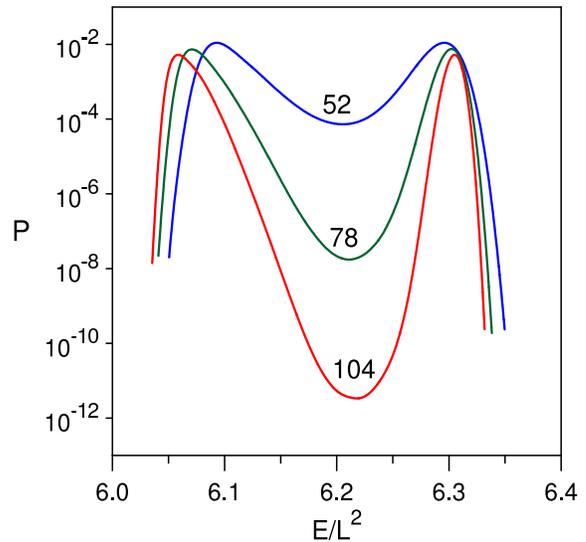}
    \caption{
  A semi--log plot of  energy distribution at transition temperature for
  3 linear sizes of system.
    }
    \label{P_Tc}
\end{figure}

\subsection {Scaling of transition temperature}

We calculated $T_c$ using fact, that the energy distribution Eq.~(
\ref{E_probab}) has two peaks of equal heights at $T=T_c$. The
results presented in Fig.\ \ref{P_Tc} demonstrate that $P(E,T_c)$
has two maxima: the first at energy $E_{-}$ of faceted phase and
the second at  energy $E_{+}$ of disordered phase. The minimum
between these peaks becomes deeper as system sizes increases. The
probability to find the system at minimum $P(E_\mathrm{min})$ is
$7\times 10^{-3}$, $2\times 10^{-6}$ and $3\times 10^{-10}$ of
$P(E_\pm)$ for system size $L$ 52, 78, and 104 respectively. From
energy distributions at $T=T_c(L)$ one can calculate the
free--energy barrier $\Delta {\cal F}(L)=-T_c(\ln
P(E_\mathrm{min},L)- \ln P(E_-,L))$ which should scale as $\Delta
{\cal F}(L)\sim L^{d-1}$ at a first order transition
\cite{kosterlitz91}. This scaling is confirmed in our calculation
because  for  $L>50$ the free--energy barrier has a linear form
fitted by $\Delta {\cal F}(L)=0.354L-13.01$.

\begin{figure}
    \includegraphics[width=8cm]{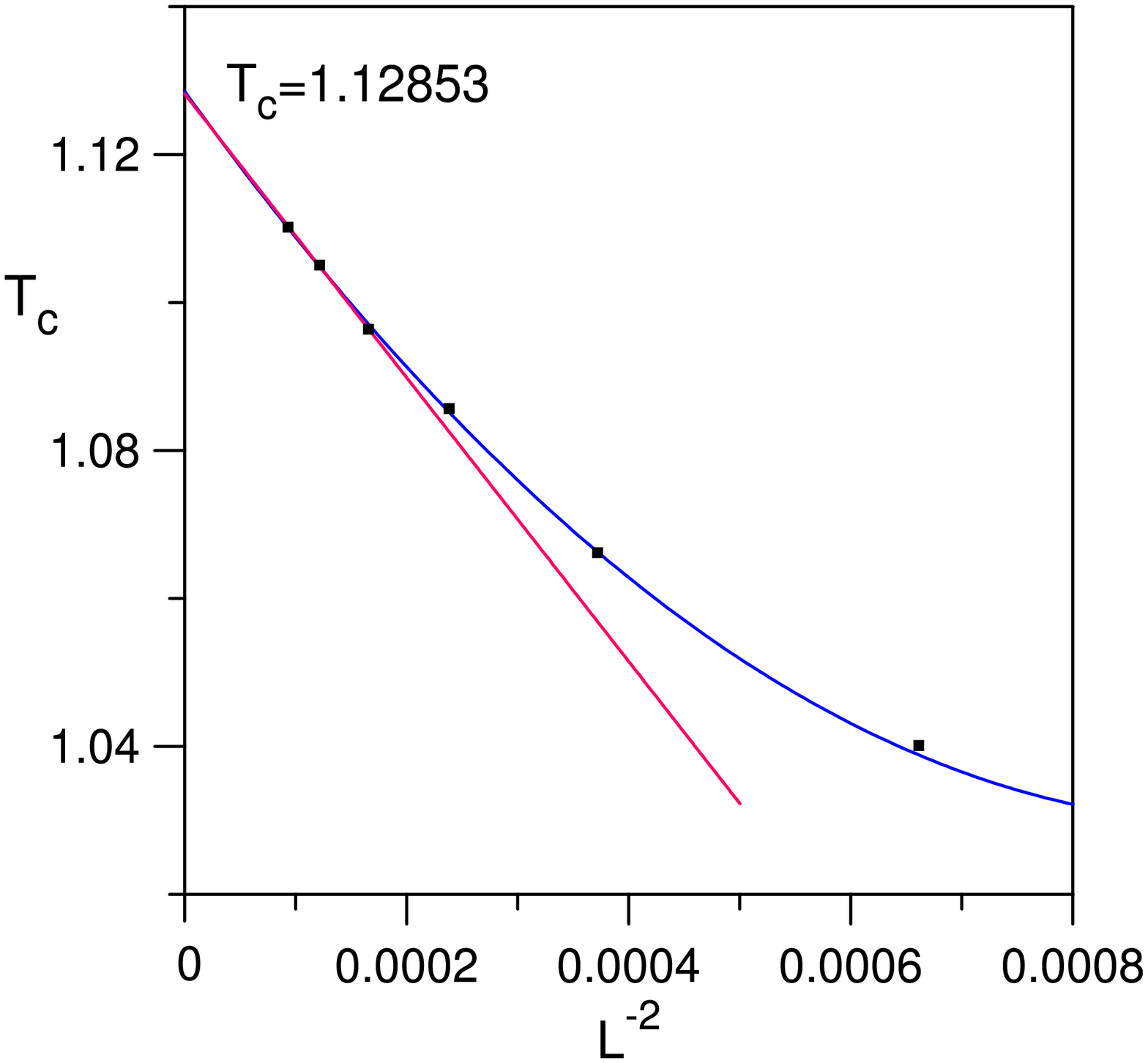}
    \caption{
 Transition temperature versus  {$L^{-2}$}. Errors are smaller than
 the symbol size. Thin and thick lines denote linear (in {$L^{-2}$})
 and non-linear fitting  of simulation result by the scaling function
 (Eq.~(\ref{scaling_Tc})).
    }
    \label{Tc_L}
\end{figure}

Having calculated $T_c(L)$ for several linear system sizes we
study scaling of $T_c(L)$  to  calculate the transition
temperature  in the limit $L\rightarrow \infty$. Our results agree
with theory of scaling at first order phase transition
\begin{equation}\label{scaling_Tc}
T_c(L)= T_c+ a_1 L^{-2}+a_2 L^{-4}
\end{equation}
The second term proportional to $L^{-4}$  is needed for system
size $L<75$ (see Fig.~\ref{Tc_L}). The transition temperature
$T_c=1.12852$ is obtained from fitting the results  with $L>50$ by
the function from Eq.~(\ref{scaling_Tc}). Similar result
$T_c=1.12829$ is obtained by applying the Burlisch --Stoer
extrapolation \cite{numrec} to $T_c(L)$. We also studied scaling
of temperature $T^s_c(L)$ at which the specific heat has a maximum
in system of linear size $L$. The differences between $T_c(L)$ and
$T^s_c(L)$ are of order $10^{-4}$ and extrapolation of $T^s_c(L)$
for $L\rightarrow \infty$ yields $T_c=1.12837$. From these two
results we estimate transition temperature for $K=-1$ as
$T_c=1.12844\pm 0.00008$.

\subsection {Scaling of specific heat}

We found that scaling of the specific heat of the system discussed
in this paper can be well described by the function of the form
used by Challa \textit{et al}. \cite {challa86} for q--state Potts
model.
\begin{equation}\label{scal_C}
C(L)=\frac{L^2Q^2(L)D}{{\left[\exp(X(L))+D \exp(X(L))\right]}^2},
\end{equation}
where
\[
Q(L)=\frac{E_{+}(L)-E_{-}(L)}{L^2T_c},
\]
and
\[
X(L)=\frac{L^2Q(L)}{2T}\left(T-T_c(L)\right ).
\]
The parameter $D$  in Eq.~(\ref{scal_C}) replaces the expression
$q(C_{-}/C_{+})^{1/2}$ for the q--state Potts model. The value of
$D=1.62$  minimizes  the sum of deviation of simulated data from
the scaling function  for $L>60$.

\begin{figure}
    \includegraphics[width=8cm]{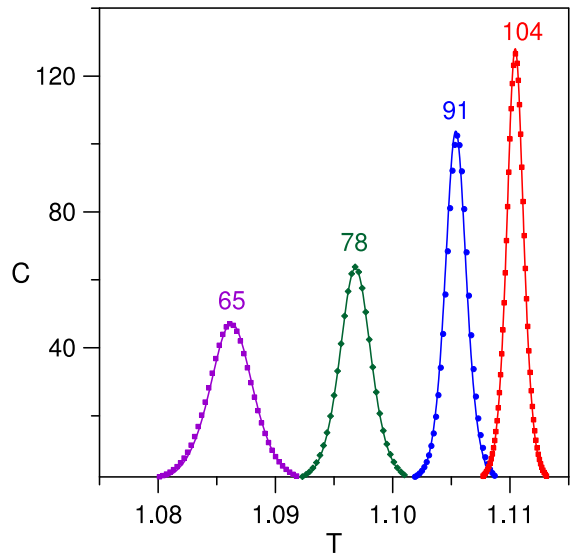}
    \caption{
  Scaling of the specific heat. Line represents the scaling function (Eq.~(\ref{scal_C})),
  symbols represent simulation data and label denotes linear system size $L$.
    }
    \label{C_L}
\end{figure}

As seen in Fig.~\ref{C_L}, the scaling function
(Eq.~(\ref{scal_C})) well describes the shape of the specific heat
near transition temperature $T_c(L)$  for $L>60$. For smaller
systems studied here ( $L<60$ ) this scaling does not apply
-- it yields incorrect position and height of specific heat
maximum.

\section{Disordered phase}\label{ScHTP}

A nature  of disordered phase is not clear. LEED experiment
results \cite{song95} suggested existence of flat phase above the
faceting temperature. On the other hand, previous  MC
simulation\cite{oleksy} indicated that this phase is not flat one
but disordered faceted phase characterized by chaotic
hill--and--valley structure which comprises  of randomly
distributed small facets mainly of \{112\} orientation. To clarify
this problem we study the size--dependence of  the mean--square
width of the surface

\begin{equation} {\delta h}^2 =
  \left<
    \frac{1}{L^2}
      \sum\limits_{j}
         {
             \left( h_j  - \bar{h}\right)^2
         }
    \right>,
\end{equation}
where sum runs over lattice sites and $\bar{h}$ is the arithmetic
average of column heights. This quantity  has been used in
investigation of roughening transition
\cite{tosat96,jaszczak97,jaszczak97ss} in SOS models
because ${\delta h}^2$ as function of a system size $L$ has
logarithmic dependence in the rough phase.
\begin{figure}
    \includegraphics[width=8cm]{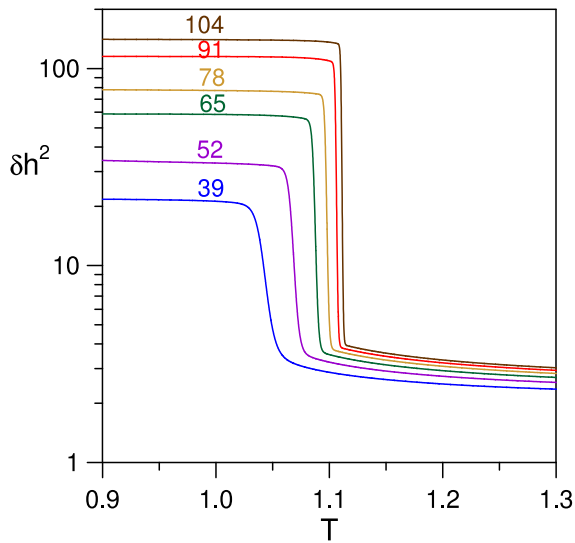}
    \caption{
 Semi--log plot of   ${\delta h}^2$ versus $T$ for several lattice sizes $L$.
    }
    \label{dhT}
\end{figure}

In order to investigate the dependence of ${\delta h}^2$ on
temperature and system size $L$ we computed the microcanonical
averages ${\delta h}^2(E)$ for each energy state in  systems with
$L<105$. The microcanonical averages ${\delta h}^2(E)$ were
calculated by performing the random walk in the energy space using
earlier computed densities of states $g(E)$  via BP method. To
assure high accuracy of  ${\delta h}^2(E)$ each state $E$ was
visited on average $10^{9}$ times.

The canonical average ${\delta h}^2(T)$ is obtained as
\begin{equation}
{\delta h}^2(T)=\frac{ \sum\limits_{E}{\delta h}^2(E)
\exp(S(E)-E/T)}{\sum\limits_{E} \exp(S(E)-E/T) }
\end{equation}
This way of calculating ${\delta h}^2(T)$ is similar to a method
of computing moments of magnetization \cite{schulz05}.

\begin{figure}
    \includegraphics[width=8cm]{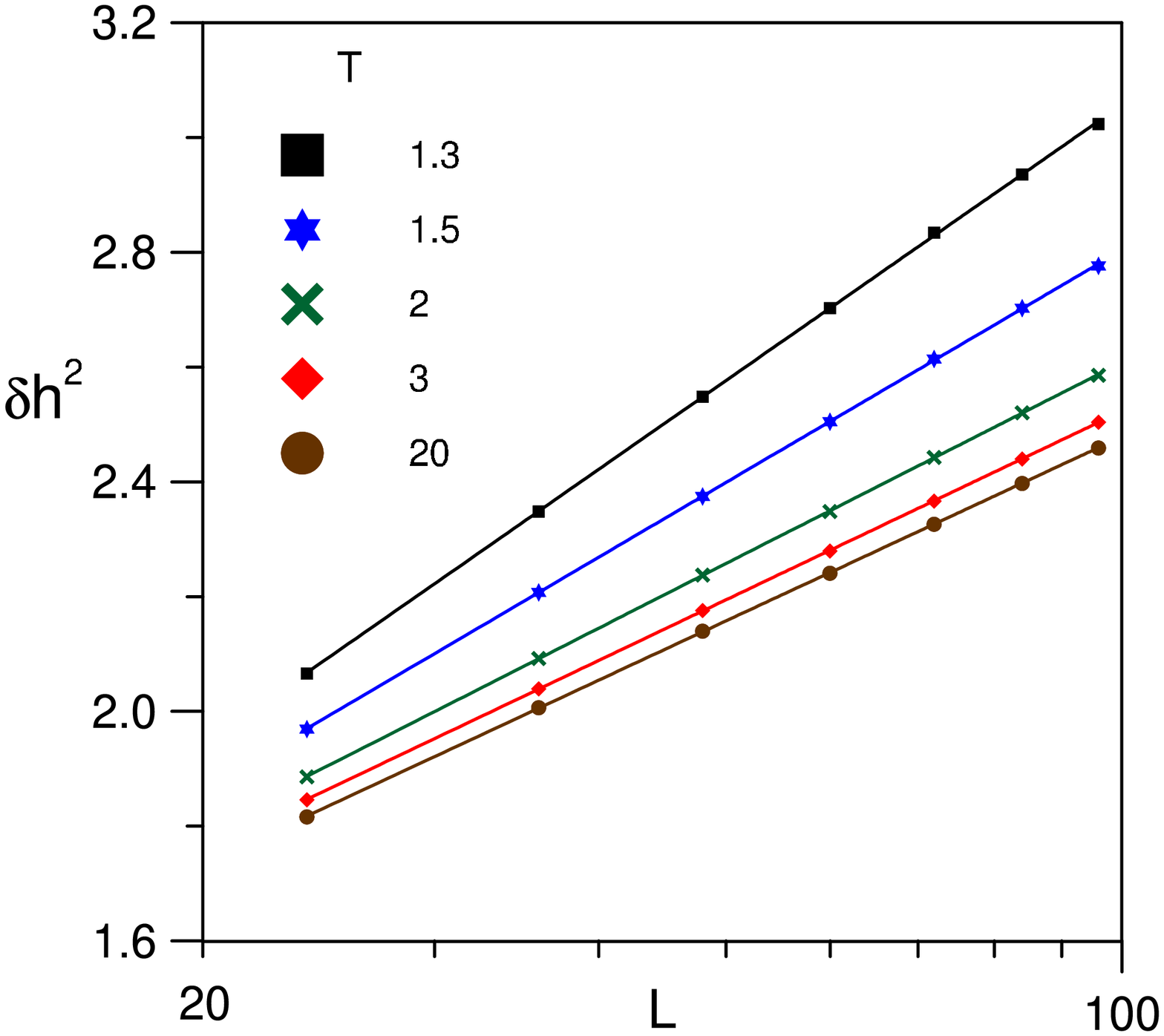}
    \caption{
 Semi--log plot of  size--dependence of ${\delta h}^2$ at several temperatures $T$.
    }
    \label{dhE}
\end{figure}

\begin{figure}
    \includegraphics[width=8cm]{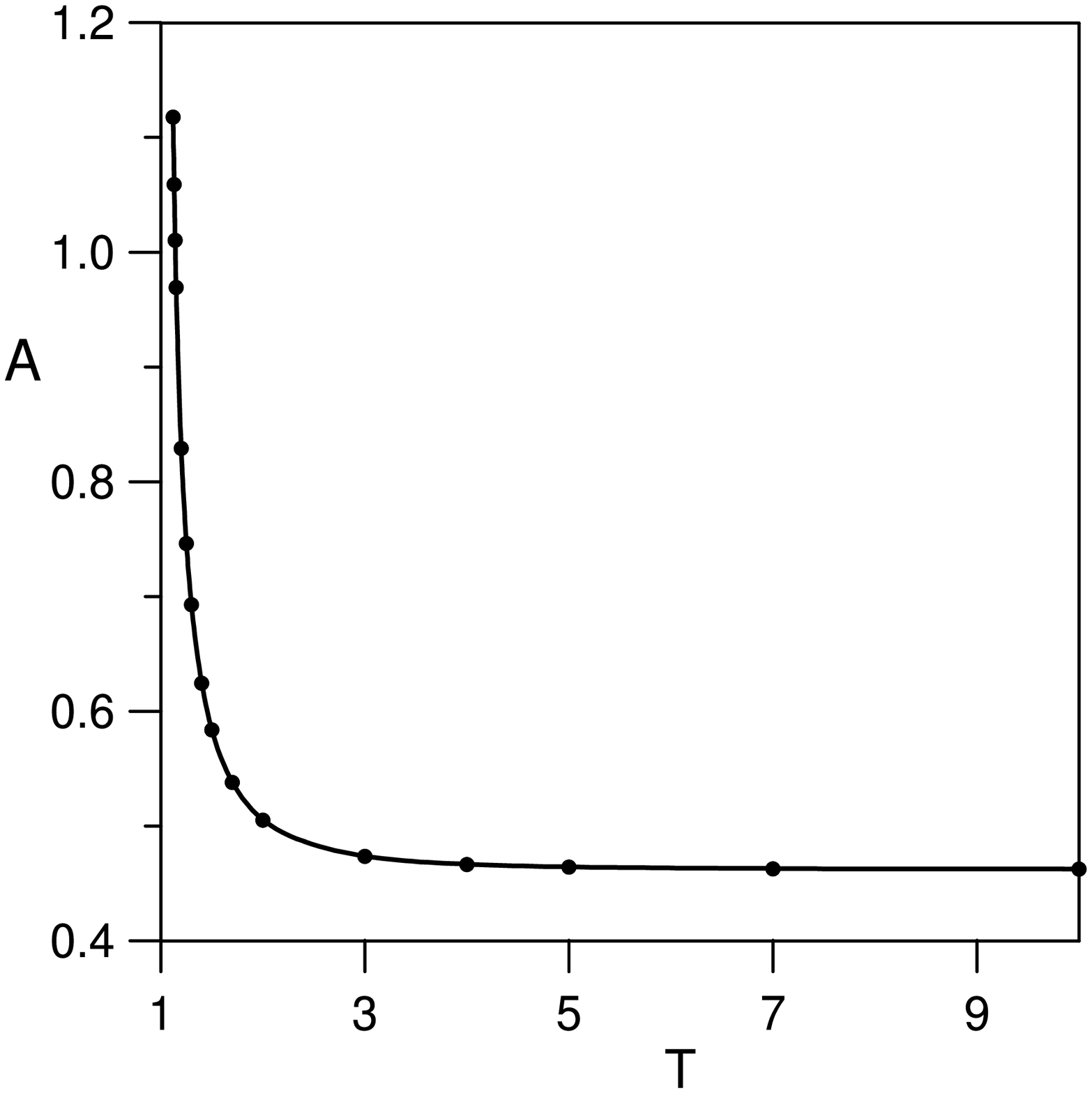}
    \caption{
 Plot of amplitude $A$ versus temperature  $T$.
    }
    \label{AE}
\end{figure}

Simulation results show (see Fig.~\ref{dhT}) that ${\delta
h}^2(T)$ is decreasing function of temperature and it rapidly
changes at transition temperature $T_c(L)$. On the other hand,
${\delta h}^2$ is increasing function of linear system size $L$.
In the high--temperature phase ${\delta h}^2$ scales
logarithmically with $L$ in the whole temperature range $T>T_c$
(see Fig.~\ref{dhE}) and results of simulations are very well
fitted by the function
\begin{equation}\label{scaling_dh2}
{\delta h}^2(T,L)= A(T)\ln L + B(T)
\end{equation}
The amplitude $A$ has the largest value at $T=T_c$ and it
decreases down to a saturation value 0.4625 for $T>T_c$ (see
Fig.~\ref{AE}).
The logarithmic dependence of the mean--square width of  the
surface on linear size of the system indicates that the disordered
phase is a rough phase. Hence, the faceting transition is the
first--order roughening phase transition. On the other hand, the
surface above $T_c$, has a disordered hill--and--valley structure
as follows from Monte Carlo simulations (see e.g. Fig.~5 in
Ref.\cite{oleksy}). Hence, this is different type of rough surface
than that observed in typical roughening transition where a
surface becomes rough by formation of steps\cite{jaszczak97ss}.

\section{Discussion}

It is demonstrated that the Wang--Landau method can be applied
to construct a phase
diagram for a  system with competitive interactions if one
performs calculations of density of states in a special
two--dimensional energy space. From such density of states one can
compute thermodynamical quantities for any value of interaction
energies. Then one can find phase transition lines and nature of
transitions. However such approach is limited to rather small
linear sizes ( in our case for $L<40$) due to huge numbers of
states occurring  in two--dimensional energy space. Because of
that, a finite--size analysis should be performed for chosen
values of interaction energies to find the order of the phase
transition, transition temperature  and other interesting
quantities.

We found that  transition from  faceted \{112\} phase to
high--temperature  phase is of first order because the energy
distribution has double--peak form at transition temperature. The
transition temperature scales as $L^{-2}$ for large linear system
size $L$, but higher order corrections are needed to the scaling
for $L<75$. Scaling of the specific heat is well described by the
function of the form used by Challa \textit{et al}. \cite
{challa86} for q--state Potts model.

In order to clarify the nature of  high--temperature phase we
calculated  the mean--square width of the surface ${\delta h}^2$.
Experimental results\cite{song95} suggest that surface becomes
flat above transition temperature. However, investigation of
surface morphology in high temperatures is not so easy. In most of
experiments, such investigations are performed after quick cooling
the sample to low temperatures. In case of adsorbate induced
faceting, attempts of  freezing of high--temperature phase were
always failed for available  cooling rates\cite{song95}.
On the other hand, the nature of  high--temperature phase can be
easy investigated in the SOS model via the Wang--Landau method.
We found that the mean--square width of the surface depends logarithmically on
the linear system size above the transition temperature. This
means that surface is rough above the faceting transition temperature.
However, this rough surface has disordered hill--and--valley structure and
differs from typical rough phase observed in roughening transition\cite{jaszczak97ss}.


\begin{thebibliography}{99}
\bibitem{song95}{K.-J.~Song, J.~C.~Lin, M.~Y.~Lai, and Y.~L.~Wang,
Surf. Sci.~327, 17 (1995).}
\bibitem{mad96}{T.~E.~Madey, J.~Guan, C.-H.~Nien, C.-Z.~Dong, H.-S.~Tao,
 and R.~A.~Campbell, Surf. Rev. Lett.~3, 1315 (1996). }
\bibitem{mad99a}{T.~E.~Madey, C.-H.~Nien, K.~Pelhos, J.~J.~Kolodziej,
 I.~M.~Abdelrehim,  and H.-S.~Tao, Surf. Sci.~438, 191 (1999).}
\bibitem{mad99b}{C.-H.~Nien, T.~E.~Madey, Y.~W.~Tai, T.~C.~Leung,
J.~G.~Che, and C.~T.~Chan, Phys. Rev. B~59, 10335 (1999).}
\bibitem{song07}{Y.-W.~Liao, L.~H.~Chen, K.~C.~Kao, C.-H.~Nien, M.-T.~Lin,
and K.-J.~Song, Phys.~Rev.~B 75, 125428 (2007). }
\bibitem{leung97} {J.~G.~Che, C.~T.~Chan, C.~H.~Kuo, and T.~C.~Leung,
Phys. Rev. Lett.~79, 4230 (1997).}
\bibitem{mad08}{T.~E.~Madey, W.~Chen, H.~Wang, P.~Kaghazchi,
 and T.~Jacob, Chem. Soc. Rev.~37, 2310 (2008).}
\bibitem{tsong01}{T.-Y.~Fu, L.-C.~Cheng, C.-H.~Nien, and T.T.~Tsong,~Phys.~Rev.~B~64, 113401
(2001).}
\bibitem{szczep05}{A.~Szczepkowicz and R.~Bryl, Phys. Rev.~B~71, 113416 (2005).}
\bibitem{szczep05b}{A.~Szczepkowicz, A.~Ciszewski, R.~Bryl, C.~Oleksy,
C.-H.~Nien, Q.~Wu, and T.~E.~Madey, Surf. Sci. 559, 55 (2005).}
\bibitem{tsong08}{H.-S.~Kuo, I.-S.~Hwang,T.-Y.~Fu, Y.-H.~Lu, C.-Y.~Lin, and T.T.~Tsong,
Appl.~Phys.~Lett. 92, 063106 (2008). }
 \bibitem{tsong09}{C.-C.~Chang, H.-S.~Kuo, I.-S.~Hwang and T.T.~Tsong, Nanotechnology 20, 115401 (2009).}
\bibitem{beijeren95}{G.~Mazzeo, E.~Carlon, and  H.~van~Beijeren,
Phys. Rev. Lett.~74, 1391 (1995). }
\bibitem{tosat96}{G.~Santoro, M.~Vendruscolo, S.~Prestipino, and  E.~Tosatti,
Phys. Rev. B~53, 13169 (1996).}
\bibitem{jaszczak97}{ D.~L.~Woodraska and J.~A.~Jaszczak, Phys. Rev. Lett.~78, 258 (1997).}
\bibitem{jaszczak97ss}{ D.~L.~Woodraska, J.~A.~Jaszczak,
Surf. Sci.~374, 319 (1997).}
\bibitem{kaseno97}{ V.~P.~Zhdanov and B.~Kaseno, Phys. Rev. B~56, R10067 (1997).}
\bibitem{oleksy}{C.~Oleksy, Surf. Sci.~549, 246 (2004).}
\bibitem{Niewiecz06}{D.~Niewieczerzal and C.~Oleksy, Surf. Sci.~600, 56 (2006). }
\bibitem{kosterlitz91}{J.~Lee and J.~M.~Kosterlitz, Phys.~Rev.~B~43, 3265 (1991). }
\bibitem{wl_0}{F.~Wang and D.~P.~Landau, Phys.~Rev.~Lett.~86, 2050 (2001).}
\bibitem{wl_1}{F.~Wang and D.~P.~Landau, Phys.~Rev.~E~64, 056101 (2001).}
\bibitem{wl_3}{S.-H.~Tsai, F.~Wang, and D.~P.~Landau, Phys.~Rev.~E~75, 061108 (2007).}
\bibitem{stampfl}{S.~Piccinin and C.~Stampfl, Phys.~Rev.~B~81, 155427  (2010).}
\bibitem{pereyra_1}{R.~E.~Belardinelli and V.~D.~Pereyra, J.~Chem.~Phys.~127, 184105 (2007).}
\bibitem{pereyra_2}{R.~E.~Belardinelli and V.~D.~Pereyra, Phys.~Rev.~E~75, 046701-1 (2007).}
\bibitem{bhatt}{C.~Zhou and R.~N.~Bhatt, Phys.~Rev.~E~72, 025701(R) (2005).}
\bibitem{numrec}{W.~H.~Press, B.~P.~Flannery, S.~A.~Teukolsky, and W.~T.~Vetterling,
Numerical Recipies. The Art of Scientific Computing, Cambridge University Press, (Cambridge, 1986).}
\bibitem{challa86}{M.~S.~S.~Challa,  D.~P.~Landau, and K.~Binder, Phys.~Rev.~B 34, 1841 (1986).}
\bibitem{schulz05}{B.~J.~Schulz and K.~Binder, Phys.~Rev.~E~71, 046705 (2005).}

\end{thebibliography}
\end{document}